\documentclass[twocolumn,prd,amsmath,superscriptaddress]{revtex4}
\usepackage{graphicx}
\usepackage{latexsym}
\usepackage{amsmath}
\usepackage{amssymb}
\usepackage{graphics}
\usepackage{color}
\usepackage{hyperref}
\usepackage{ifthen}
\usepackage{bm}

\definecolor{darkgreen}{RGB}{0,180,0}

\newcommand{\bmat}{\xi}  
\newcommand{\dVeff}{V_{\mathrm{eff},\phi}}
\newcommand{\DVloop}{\Delta V_\mathrm{1-loop}}
\newcommand{\meff}{m_\phi}       
\newcommand{\mmax}{m_\mathrm{max}}
\newcommand{\Mpl}{M_\mathrm{Pl}} 
\newcommand{\phibulk}{\phi_\mathrm{m}}   
\newcommand{\rholab}{\rho_\mathrm{lab}}
\newcommand{\rhomat}{\rho} 
\newcommand{\Veff}{V_\mathrm{eff}} 
\newcommand{\Vloop}{V_\mathrm{1-loop}}

\begin{document}

\title{Quantum Stability of Chameleon Field Theories}
\author{Amol Upadhye}
\affiliation{Argonne National Laboratory, 9700 S. Cass Ave., Lemont, IL 60439}%

\author{Wayne Hu}
\affiliation{Kavli Institute for Cosmological Physics, Department of Astronomy \& Astrophysics, University of Chicago, Chicago, IL, 60637
}%

\author{Justin Khoury}
\affiliation{Center for Particle Cosmology, Department of Physics and Astronomy, University of Pennsylvania, Philadelphia, PA 19104
}%

\date{\today}

\begin{abstract}
Chameleon scalar fields are dark energy candidates which suppress fifth forces in high density regions of the universe by becoming massive.  We consider chameleon models as effective field theories and estimate quantum corrections to their potentials.  Requiring that quantum corrections be small, so as to allow reliable predictions of fifth forces, leads to an upper bound $m < 0.0073 (\rhomat / 10~{\rm g\, cm}^{-3})^{1/3}$eV for gravitational strength coupling whereas fifth force experiments place a lower bound of $m>0.0042$\,eV.
An improvement of less than a factor of two in the range of fifth force experiments could test all classical chameleon field theories whose quantum corrections 
are well-controlled and couple to matter with nearly gravitational strength regardless of the specific form of the chameleon potential.  
\end{abstract}

\maketitle



{\em Introduction.--} Cosmic acceleration, discovered over a decade ago, is the great mystery of modern cosmology.  Since the simplest model, a cosmological constant, offers no clues as to the smallness of the acceleration or to its recent onset, the search for other explanations for the acceleration is an active area of research.  The next simplest models typically involve a scalar field, which is likely to couple to matter in the absence of a symmetry forbidding such a coupling~\cite{Frieman_Hill_Stebbins_Waga_1995,Carroll:1998zi}.  Gravitational-strength fifth forces have been excluded over a large range of length scales~\cite{Kapner_etal_2007,Adelberger_etal_2009}, so a viable scalar theory must contain a mechanism for screening such forces locally.  Chameleon~\cite{Khoury_Weltman_2004a,Khoury_Weltman_2004b,Brax_etal_2004} fields become massive in high density regions of the universe, pushing fifth forces to length scales below the bounds of current experiments. Symmetron~\cite{Hinterbichler_Khoury_2010,Olive:2007aj,Pietroni:2005pv} and Galileon~\cite{Deffayet:2001uk,Nicolis:2008in} fields lower their effective couplings to matter in high density regions through symmetry restoration and higher-derivative interactions, respectively.

Although progress has been made toward embedding chameleon models in more fundamental theories~\cite{Brax_Martin_2007,Hinterbichler_Khoury_Nastase_2011}, for now they are 
best treated as effective field theories, valid only below a certain potential-dependent cutoff energy scale. Above the cutoff, quantum corrections to the potential make predictions
of fifth forces unreliable. 
In this paper, we estimate the one-loop Coleman-Weinberg correction to the potential, arising from chameleons running in the loop.
Demanding that quantum corrections remain small compared to the classical potential, we find that the resulting ``classical'' chameleon theories cannot acquire masses larger than 
$\meff \sim (\bmat \rhomat / \Mpl)^{1/3}$ at a density $\rho$ and dimensionless matter coupling $\bmat$.  (Here $\Mpl = (8\pi G)^{-1/2}$ is the reduced Planck mass and $\hbar=c=1$ throughout.)
Viable chameleons must therefore tiptoe between being heavy enough to avoid fifth
force constraints and remaining light enough to keep quantum corrections under control. We will find that independently of the specific form for the
self-interactions, there is tension between keeping quantum corrections under control and satisfying laboratory constraints on fifth forces.

Numerically, our upper bound can be expressed as $\meff< 0.0073(\bmat\rhomat/10~\text{g cm}^{-3})^{1/3}$\,eV.  This energy scale is interesting for dark energy models and is also accessible to upcoming fifth force experiments.  Of course, there is no requirement for Nature  to choose a model which remains a valid effective theory out to scales accessible to experiments.   However, these classical theories are the only known chameleon models with predictive power there, so our analysis can offer guidance as to the regions of the theory parameter space on which future experiments should focus.


\smallskip
{\em Chameleon fields.--}
Consider a chameleon scalar field with equation of motion $\Box \phi = \dVeff$, where the effective potential is~\cite{Khoury_Weltman_2004a,Khoury_Weltman_2004b,Brax_etal_2004}
\begin{eqnarray}
\Veff(\phi,\vec x)
&=&
V(\phi) + \frac{\bmat \rhomat(\vec x) \phi}{\Mpl}\,.
\label{e:Veff}
\end{eqnarray}
Here $\bmat$ is a dimensionless coupling to the matter density $\rhomat$, and the bare potential $V(\phi)$ is a function of the field alone.  The matter coupling is a linearization, valid for $|\phi| \ll  \Mpl/\bmat$, of the general scalar-tensor form $-\exp(\bmat\phi/\Mpl) T^\mu_{\;\mu}$, where the trace of the stress tensor reduces to $T^\mu_{\;\mu} \approx -\rhomat$ for nonrelativistic matter.

The effective mass depends on field value $\meff^2 = V''(\phi)$.  
Inside a sufficiently large bulk of constant matter density, the field settles to
its equilibrium value of  
\begin{equation}
V'(\phibulk) =- \bmat\rhomat/\Mpl \,.
\label{eqn:phirho}
\end{equation}
The potential $V(\phi)$ is chosen so that $\meff(\phibulk(\rhomat))$ increases with $\rhomat$; thus the range of the fifth force mediated by $\phi$ shrinks with increasing $\rhomat$.  By becoming more massive in higher-density regions such as the laboratory, the chameleon field can ``hide'' from small-scale fifth force tests including~\cite{Kapner_etal_2007}. 



\smallskip
{\em Quantum corrections.--}
Both the self-interaction and the matter coupling in (\ref{e:Veff}) will give rise to quantum corrections to a chameleon theory.  Matter loops will of course generate large radiative corrections to the chameleon mass. This is the origin of the hierarchy problem, and we have nothing new to add here. In this work we therefore focus on quantum corrections due to $\phi$ loops, and treat the matter as an external source. In particular, we treat the matter coupling $\bmat$ as a free parameter, and focus on scalar quantum corrections to the potential. Even with these optimistic assumptions about the matter, we will find that quantum corrections from $\phi$ alone impose stringent constraints on the potential.

The one-loop Coleman-Weinberg correction to the classical potential $V(\phi)$, neglecting spatial variations in the field, is given by
\begin{equation}
\DVloop(\phi)
=
\frac{\meff^4(\phi)}{64\pi^2}
\ln \left(\frac{\meff^2(\phi)}{\mu_0^2}\right)\,,
\label{e:DVloop}
\end{equation}
where $\mu_0$ is some arbitrary mass scale.  The one-loop corrected potential is then $\Vloop = V + \DVloop$.  Even if we choose the mass scale to eliminate the correction 
at some density $\rho_0$, $\mu_0 = \meff(\phibulk(\rho_0))$, the fact that $\phi$ is a chameleon field where the mass runs with field value will imply corrections at other densities. 
When the one-loop corrections become as large as the tree-level terms,  there is no reason to believe that higher-order loop corrections will not also be significant. Thus we use the corrections arising from $\DVloop$ as estimates of the quantum uncertainty in the chameleon model.  A given classical chameleon model is predictive only if these quantum corrections are small at densities of interest.

Since  $\DVloop \sim \meff^4$, we can immediately see that quantum corrections can present problems for chameleon theories.  Chameleon screening of fifth forces operates by increasing $\meff$, so quantum corrections must become important above some effective mass.   
On the other hand, laboratory measurements place a lower bound on the effective mass 
leading to tension between a model's classical predictivity and the predictions that it makes.

Specifically, for the chameleon mechanism to be classically predictive we require
both $\DVloop'/V'$ and $\DVloop''/V''$ to be small across the field range of interest.
The former sets the field value $\phibulk$ and the latter sets the effective mass at that value.
Note that large one-loop corrections do not necessarily imply the breakdown of effective field theory.  An unnatural theory could have large one-loop terms, with higher-order terms small, although this is not the case for the power law models considered in later sections.  

\smallskip
{\em 1-loop bound on mass.--}
While we can always evaluate the loop bounds in the previous section and compare them with laboratory bounds  for any given chameleon potential $V(\phi)$, it is useful
to phrase the main physical content of the bound in a model independent manner.

Classicality imposes a limit on the effective mass which a chameleon field may attain in an experiment, which depends on the density $\rholab$ of the experimental apparatus.  
As a simple estimate, we set the log term in (\ref{e:DVloop}) to unity
so that our loop criteria becomes
\begin{eqnarray}
\left|\frac{\DVloop'}{V'}\right| 
&\approx& 
\left|
\frac{\Mpl}{\bmat \rhomat} \frac{(\meff^4)'}{64\pi^2}
\right|  
<
\epsilon \,;
\nonumber
\\
\left|\frac{\DVloop''}{V''}\right| 
&\approx&   
\left|
\frac{(\meff^4)''}{64\pi^2 \meff^2}
\right| 
< \epsilon\,,
\label{eqn:epsilon}
\end{eqnarray}
where $\epsilon$ should not exceed unity. Using 
Eq.~(\ref{eqn:phirho}) 
the field derivatives can be replaced with density derivatives through
${\rm d}\phibulk/{\rm d}\rho = -\bmat \Mpl^{-1} \meff(\phibulk)^{-2} $
to obtain
\begin{equation}
\frac{1}{\rhomat}
\frac{{\rm d} \meff^6}{{\rm d}\rhomat}
\, , \,
\left|
\frac{{\rm d}^2\meff^6}{{\rm d}\rhomat^2}
\right|
\leq
\frac{96\pi^2 \bmat^2}{\Mpl^2} \epsilon\,.
\end{equation}
At the density $\rholab$ these imply
\begin{equation}
\meff
\leq
\left(\frac{48\pi^2 \bmat^2 \rholab^2 \epsilon}{\Mpl^2}\right)^\frac{1}{6}
=
0.0073 \left( \frac{\bmat \rholab}{10~{\rm g\,  cm}^{-3}} \right)^\frac{1}{3} \epsilon^\frac{1}{6}\,  {\rm eV}\,.
\label{e:mbound}
\end{equation}
For $\bmat\sim \epsilon \sim 1$ and typical densities this mass scale is close to the dark energy scale of $\rho_\Lambda^{1/4} = 0.0024$\,eV. This results from the numerical coincidence that $(\rholab/\Mpl)^{4/3} \sim \rho_\Lambda$. 
Importantly, the dependence on $\epsilon$ is weak and the Compton wavelength corresponding to this maximum mass, $0.027 (\bmat \rholab/10~{\rm g\, cm}^{-3})^{-1/3} \epsilon^{-1/6}$~mm, is comparable to the length scales probed by the smallest-scale torsion pendulum experiments.   
Given this weak dependence, henceforth we set $\epsilon = 1$, the largest value at which order-unity predictions of fifth forces could reasonably be trusted.

\begin{figure}[tb]
\begin{center}
\includegraphics[angle=270,width=3.3in]{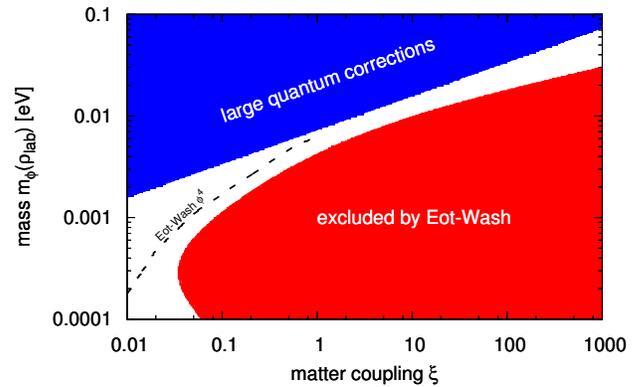}
\caption{Model-independent constraints on chameleon fields in the $\bmat$,~$\meff$ plane with $\rholab = 10$~g/cm$^3$.  Shaded regions show loop bounds from (\ref{e:mbound}) and experimental constraints from E\"ot-Wash~\cite{Kapner_etal_2007}.  The dashed curve shows the direct bound on the $\phi^4$ model  for $\bmat<1$~\cite{Adelberger_etal_2007}, converted to $\meff$.  Our bound is  conservative in that it allows slightly lower values of $\meff$. 
  \label{f:mloop_and_eotwash}}
\end{center}
\end{figure}


\smallskip{\em  Tension with Laboratory Bounds.--}
Torsion pendulum experiments such as E\"ot-Wash~\cite{Kapner_etal_2007} exclude fifth forces due to Yukawa scalars with constant masses $m$ over a region of the $\bmat$,~$m$ parameter space. Let $\mmax$ be the maximum mass of a given chameleon model in a fifth force experiment.  The chameleon-mediated fifth force should be bounded from below by the force of the Yukawa scalar with mass $m = \mmax$.  This is because the mass of the chameleon is lighter than $\mmax$ in the lower-density regions of the experiment, so the range of its fifth force is larger.  Thus, approximating the chameleon's fifth force by the Yukawa force will lead to a conservative constraint on the chameleon; we refer to this as the maximum-mass approximation.  We quantify how much bounds are improved by a direct calculation for
specific potentials below.  Note that the maximum-mass approximation is used to place a minimum mass bound on $m_\phi$.

We show this 
 E\"ot-Wash  constraint~\cite{Kapner_etal_2007} on the minimum mass in
 Fig.~\ref{f:mloop_and_eotwash}.   We compare this to the maximum mass from the loop bound at the relevant density of  $\rholab = 10$~g/cm$^3$, working in the maximum-mass approximation.
  The tension between these two bounds is evident, especially near $\bmat = 1$.  A significant, but feasible, improvement in E\"ot-Wash constraints over the next several years of less than a factor of 2 in the Yukawa mass or fifth force range could eliminate all chameleon fields around $\bmat=1$ whose quantum corrections are well-controlled.


\begin{figure}[tb]
\begin{center}
\includegraphics[width=2.3in]{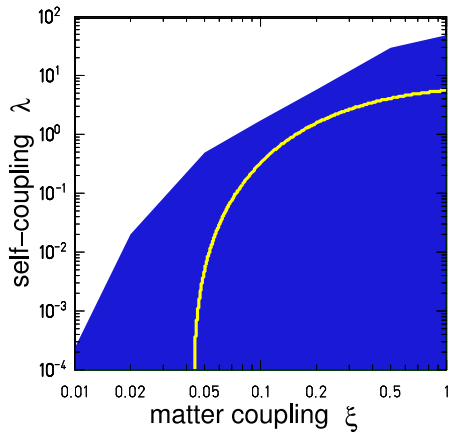}
\caption{Constraints on chameleon models with $V = \lambda\phi^4/4!$.  The shaded region shows the models excluded by~\cite{Adelberger_etal_2007} while the curve shows the weaker constraints resulting from the maximum-mass approximation.}
\label{f:constraints}
\end{center}
\end{figure}

\smallskip{\em Model Constraints.--}
Our maximum-mass approximation yields conservative but model-independent bounds on chameleon models.     In the context of particular models, our approximations can be checked
against direct computation.

In addition to constant-mass scalar theories,  the E\"ot-Wash experiment has also constrained chameleon theories with $V(\phi) = \lambda\phi^4/4!$ and $\bmat<1$~\cite{Adelberger_etal_2007}.
The $\phi^4$ theory is also special in that the loop bound (\ref{e:mbound}) is independent of $\rhomat$ and $\bmat$: since $\meff 
=\lambda^{1/6}(3\bmat \rhomat/\Mpl)^{1/3}$ in this case, it follows that $\lambda < 32\pi^2 \epsilon / 3 \approx 105 \epsilon$. 
In Fig.~\ref{f:mloop_and_eotwash}, we convert their constraints on $\lambda$, shown in Fig.~\ref{f:constraints}, to a bound on $\meff$.   

As expected the direct $\lambda$ bound rules out slightly more of the $\meff$ space than
our mass bound for gravitational strength $\bmat$ but there is still an allowed region which
satisfies both the loop and the laboratory bound.  
As also shown in Fig.~\ref{f:constraints}, the impact of our approximation on the $\lambda-\bmat$ parameter space is more pronounced since $\lambda \propto \meff^6$ but correspondingly the loop-compatible range appears larger and includes
all of the space shown.   Nonetheless it is the mass  that is more closely related
to the experimental observables and even with our conservative assumptions a factor of
2 there would close the $\bmat\sim1$ window entirely in this model.

We in fact expect our constraints to be conservative for generic chameleon models.  
To see this, consider the case of power law potentials 
\begin{equation}
V(\phi) 
=
\kappa M_\Lambda^{4-n} |\phi|^n\,,
\end{equation}
where the arbitrary mass scale $M_\Lambda$ is suggestively set to the
dark energy scale $M_\Lambda = 0.0024$\,eV, thereby making $\kappa$  a dimensionless constant.
To have a chameleon model with a bounded potential requires $n <0$ or $n>2$. 
Note that our bounds would be unchanged by adding in a constant $M_\Lambda^4$ or
a slowly varying piece to the potential that plays the role of a cosmological constant.

To model the experimental set up, consider 
a constant-density planar slab
surrounded by vacuum: $\rho(x) = \rholab$ for $x \leq 0$ and $\rho(x) = 0$ for positive $x$.
Using the exact solutions of Refs.~\cite{Upadhye_Gubser_Khoury_2006,Brax_etal_2007} for $V(\phi) \propto |\phi|^n$ in the vacuum $x>0$,
\begin{equation}
\phi(x)
=
\frac{\left(1-\frac{1}{n}\right)\phibulk(\rholab)}
     {\left( 1 + \sqrt{\frac{1}{2}\left| \frac{(n-2)^2(n-1)^{n-3}}{n^{n-1}} \right|} \meff(\rholab) x   \right)^\frac{2}{n-2}}\,,
\end{equation}
we  can evaluate the acceleration $a_\phi = -(\bmat / \Mpl){\rm d}\phi/{\rm d}x$ of a test particle.  A Yukawa scalar $\varphi$ with $m = \mmax \equiv \meff(\rholab)$ and the same matter coupling $\bmat$ will cause an acceleration $a_\varphi = -(\bmat/\Mpl){\rm d}\varphi/{\rm d}x$ with 
\begin{equation}
\varphi(x) = -\frac{\bmat \rholab}{2\mmax^2\Mpl} \exp(-\mmax x)\,.
\end{equation}
  Direct comparison shows that $|a_\phi| \geq |a_\varphi|$ at $x=0$, and $|a_\phi|$ decreases more slowly than $|a_\varphi|$ for all $x \geq 0$.  Thus $|a_\phi| \geq |a_\varphi|$ everywhere.  To generalize, since $\meff < \mmax$ outside the highest-density part of the experiment, the fifth force due to a chameleon falls off more slowly with distance than that due to a Yukawa scalar with $m=\mmax$.  The chameleon force is therefore larger and easier to exclude.

\begin{figure}[tb]
\begin{center}
\includegraphics[width=3.25in]{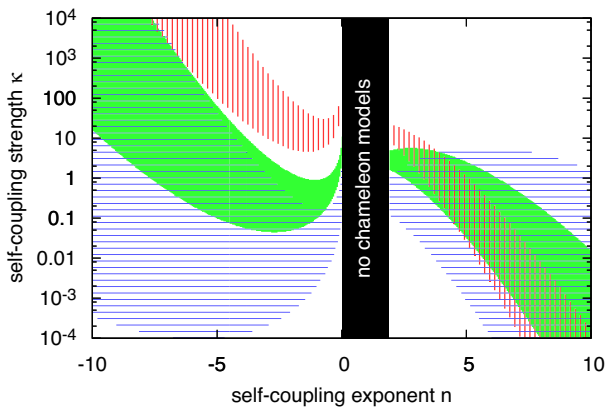}
\caption{Allowed classical chameleon models with power law potentials $V(\phi) =\kappa M_\Lambda^{4-n} |\phi|^n$, with $M_\Lambda = 0.0024$\,eV.
Blue (horizontal hatched), green (solid) and red (vertical hatched)  regions show models with $\bmat = 0.1$, $1$, and $10$, respectively, which satisfy (\ref{e:mbound}) and are consistent with~\cite{Kapner_etal_2007} in the maximum-mass approximation.
}
\label{f:models}
\end{center}
\end{figure}

The Yukawa mass limits can then be converted into conservative constraints
on the parameters of the power law potentials.
Figure~\ref{f:models}~ shows models which are consistent with the data~\cite{Kapner_etal_2007} in the maximum-mass approximation and whose quantum corrections satisfy~(\ref{e:mbound}) for various $\bmat$.  
Although one can always find allowed models by tuning $\bmat$ to sufficiently small values, couplings of gravitational strength $\bmat \sim 1$ and higher are the most interesting for chameleon theories.  

Once again, the tension between loop corrections, which impose an upper bound on $n \kappa$, and fifth force constraints, which impose a lower bound, is evident from the figure.  A modest improvement in experimental constraints could rule out all allowed models for some range of $\bmat$.  Submillimeter fifth force constraints from much denser environments, such as the $150$~g/cm${^3}$ solar center, could also rule out a substantial fraction of the models in Fig.~\ref{f:models}, though we do not have a specific probe in mind.

\smallskip{\em Conclusions.--}
We have shown that keeping quantum corrections to chameleon theories under control imposes a density-dependent upper limit on the chameleon mass which is in tension with laboratory bounds on small-scale fifth forces.  This tension can be quantified in a general, model-independent way by approximating the chameleon field by a Yukawa scalar whose constant mass equals the maximum mass of the chameleon in the experiment.  Even in this conservative approximation, only a small range of viable predictive models remains for couplings around the gravitational strength, $\bmat \sim 1$, which could be excluded by a factor-of-two improvement in bounds on the range of the fifth force.

Such an improvement would test all such chameleon models, regardless of the form for their
self-interaction.   These models include scalar-tensor theories such as the $f(R)$ model
where $\bmat = 1/\sqrt{6}$.   
Likewise they include other dark-energy motivated models where the dimensionful parameter characterizing the self-interaction is set to the dark energy scale.  

In dark-energy motivated models, the chameleon may still be invoked at lower densities, e.g.~to provide cosmological range forces which are sufficiently suppressed in the Solar system.   At these lower densities, the loop bound is relatively easier to satisfy, e.g.~at the background matter density the range $\meff^{-1} > 4 \times 10^5\xi^{-1/3}$m, allowing fifth forces on cosmological scales.  However such models would  no longer be valid effective
field theories at laboratory densities and hence would lose some of their predictive power.


\smallskip{\em Acknowledgments.--}  We are grateful to Peter Adshead, Claudia de Rham, Salman Habib, Kurt Hinterbichler, Andrew Tolley and Mark Wyman for helpful discussions. 
AU was
supported by the U.S. Department of Energy, Basic Energy Sciences, Office of
Science, under contract No. DE-AC02-06CH11357.
WH was supported by Kavli Institute for Cosmological Physics at the University of Chicago through grants NSF PHY-0114422 and NSF PHY-0551142  and an endowment from the Kavli Foundation and its founder Fred Kavli,  by U.S.~Dept.\ of Energy contract DE-FG02-90ER-40560 and the David and Lucile Packard Foundation. JK is supported in part by NASA ATP grant NNX11AI95G, and the Alfred P. Sloan Foundation. 

\small{
The submitted manuscript has been created by
UChicago Argonne, LLC, Operator of Argonne
National Laboratory (``Argonne''). Argonne, a
U.S. Department of Energy Office of Science laboratory,
is operated under Contract No. DE-AC02-
06CH11357. The U.S. Government retains for itself,
and others acting on its behalf, a paid-up
nonexclusive, irrevocable worldwide license in said
article to reproduce, prepare derivative works, distribute
copies to the public, and perform publicly
and display publicly, by or on behalf of the Government.
}

\bibliographystyle{unsrt}

\bibliography{chameleon}

\end{document}